\documentclass[twocolumn,pre,showpacs,groupedaddress,preprintnumbers,amsmath,amssymb,floatfix]{revtex4}

\usepackage{graphicx}
\usepackage{dcolumn}

\begin{document}

\title{Softening of the stiffness of bottlebrush polymers by mutual interaction}

\author{S. Bolisetty$^a$, C. Airaud$^a$, Y. Xu$^b$, A. H. E. M\"uller$^b$, L. Harnau$^c$, S. Rosenfeldt$^a$, P. Lindner$^d$,
and M. Ballauff$^a$} 
\email[]{E-mail: Matthias.Ballauff@uni-bayreuth.de; harnau@fluids.mpi-stuttgart.mpg.de} 
\affiliation{$^a$Physikalische Chemie I, University of Bayreuth, 
D-95440 Bayreuth, Germany\\ 
$^b$Makromolekulare Chemie II, University of Bayreuth, 
D-95440 Bayreuth, Germany\\  
$^c$Max-Planck-Institut f\"ur Metallforschung,  
Heisenbergstr.\ 3, D-70569 Stuttgart, Germany, 
\\and Institut f\"ur Theoretische und Angewandte Physik, 
Universit\"at Stuttgart, Pfaffenwaldring 57, D-70569 Stuttgart, Germany\\
$^d$Institut Laue-Langevin, B. P. 156X, 38042 Grenoble CEDEX 9, France}

\date{\today}

\begin{abstract}
We study bottlebrush macromolecules in a good solvent by small-angle neutron 
scattering (SANS), static light scattering (SLS), and dynamic light scattering 
(DLS). These polymers consist of 
a linear backbone to which long side chains are chemically grafted. The backbone 
contains about 1600 monomer units (weight average) and every second monomer unit 
carries side-chains with ca.~60 monomer units. 
The SLS- and SANS data extrapolated to infinite dilution lead to the form factor 
of the polymer that can be described in terms of a worm-like chain with a contour 
length of \mbox{380 nm} and a persistence length of 17.5 nm. An analysis of the 
DLS data confirm these model parameters. The scattering intensities 
taken at finite concentration can be modeled using the polymer reference 
interaction site model. It reveals a softening of the bottlebrush polymers 
caused by their mutual interaction. We demonstrate that the persistence decreases 
from 17.5 nm down to 5 nm upon increasing the concentration from dilute solution 
to the highest concentration (\mbox{40.59 g/l}) under consideration. The observed softening 
of the chains is comparable to the theoretically predicted decrease of the 
electrostatic persistence length of linear polyelectrolyte chains at finite 
concentrations.
\end{abstract}
\maketitle

If polymeric side chains are grafted to a flexible or rigid polymer backbone, a 
cylindrical bottlebrush polymer results \cite{Win:94,Win:95,Ger:99,Zha:05,Rat:05,Rat:06,Zha:06}. 
The main feature of these polymers is a marked stiffening of the main chains 
(see, e.g., the discussion in Refs.~\cite{Rat:05,Rat:06,Zha:06}). It has been 
demonstrated theoretically and by computer simulations that this stiffening 
is due to a balance of the repulsive forces originating from a steric overcrowding 
of the side chains and the entropic restoring force of the main chain \cite{Theory}. 
The analysis 
of bottlebrush polymers by small-angle neutron scattering (SANS), small-angle 
\mbox{X-ray} scattering (SAXS) and static light scattering (SLS) in dilute solution 
has supported this picture by showing that these macromolecules exhibit a worm-like 
conformation \cite{Rat:05,Rat:06,Zha:06}. However, up to now most studies 
on bottlebrush polymers in solution have focused on the dilute regime and
{\it conformational ideality} has been assumed. That is, the intramolecular pair 
correlations are presumed to be independent of polymer concentration and can 
be computed based on a chain model that only accounts for intramolecular interactions 
between monomers. However, this assumption can fail upon increasing the polymer concentration 
because the polymers begin to interpenetrate leading to a medium-induced interaction
between two monomers of individual polymers. As a result the persistence length is expected to
decrease with increasing polymer concentration in the semidilute solution regime.
Such concentration-dependent conformational changes of chain molecules 
have been investigated theoretically 
for semidilute solutions of bottlebrush polymers \cite{bori:87},
dense polymer solutions and melts 
(see, e.g., \cite{schw:92,yeth:92,mele:93,schw:97,sung:05}), 
and semiflexible chain polyelectrolyte solutions
(see, e.g., \cite{stev:93,stev:95,yeth:97,yeth:98,shew:99,shew:00,hofm:03}).
Here we present the first systematic experimental and theoretical study of 
concentration-dependent conformational changes of bottlebrush polymers which elucidates 
the importance of the medium-induced interaction on soft materials such as polymers. 
We demonstrate that mutual interaction between the bottlebrush 
polymers leads to a significant reduction of their stiffness in solution.\\
\begin{figure}[t!]
\includegraphics[ width=5.8cm,clip]{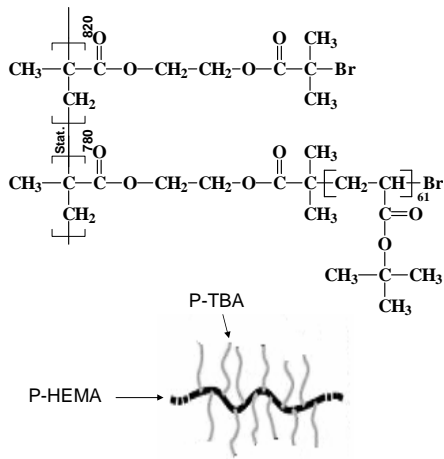}
\vspace*{-0.2cm}
\caption{Chemical structure of the investigated bottlebrush polymer consisting 
of a poly(2-hydroxyethylmethacrylate) (p-HEMA) backbone and poly(t-butyl acrylate) 
(p-TBA) side chains. The repeating units carrying the side chains alternate statistically 
with unsubstituted repeating units. The average number of repeating units per side 
chain is 61.}
\label{fig1}
\end{figure}
Figure \ref{fig1} displays the repeating unit of the polymer under consideration. 
This polymer has been synthesized by a "grafting from" method and composed of 
poly(2-hydroxyethylmethacrylate) backbone grafted with poly(t-butyl acrylate) chains. 
Details of the synthesis and the characterization have been reported in 
Ref.~\cite{Zha:03}. SANS measurements of dilute solutions of the bottlebrush polymer in deuterated tetrahydrofurane (THF) were performed at the beamline D11 of the 
Institut Laue-Langevin in Grenoble, France. The incoherent contribution to the measured 
intensities has been determined at the highest scattering angles and subtracted in order 
to obtain the coherent part. In all cases absolute intensities have been obtained. 
Details of the data evaluation may be found in Ref.~\cite{rose:06,Li:05}.

\begin{figure}[t!]
\includegraphics[width=6.2cm,clip]{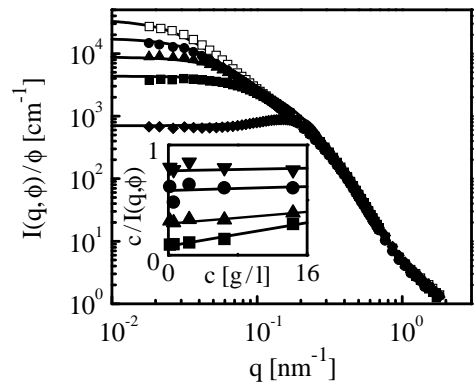}
\vspace*{-0.1cm}
\caption{Absolute scattering intensities $I(q,\phi)$ of bottlebrush polymer solutions
normalized to their volume fraction $\phi$. The open squares and the upper line
represent the intensity extrapolated to vanishing concentration and the calculated 
form factor of a worm-like chain, respectively. The solid symbols denote the measured 
intensities for four bottlebrush polymer concentrations 
(circles, $c = 2.40$ g/l; triangles, $c = 6.45$ g/l; squares, $c = 14.35$ g/l; 
diamonds, $c = 40.59$ g/l). The four lower lines represent the corresponding
intensities as obtained from the PRISM integral equation theory 
[Eqs.~(\ref{eq3}) and (\ref{eq4})] and taking into 
account the softening of the bottlebrush polymers (see Figs.~\ref{fig3} (a) and (b)).
The inset shows the extrapolation of the measured intensities to vanishing 
concentration according to Eqs.~(\ref{S(q)}) and (\ref{Bapp}) for four 
scattering vectors: down triangles, $q=0.15$ nm$^{-1}$;
circles, $q=0.12$ nm$^{-1}$; up triangles, $q=0.08$ nm$^{-1}$;
squares, $q=0.04$ nm$^{-1}$.}
\label{fig2}
\end{figure}

Without loss of generality, the measured scattering intensity, $I(q,\phi)$, as a
function of the magnitude of the scattering vector $q=|\vec q\,|$ and 
the volume fraction of the solute $\phi$ can be rendered 
as the product of a form factor $P(q)$ and a structure factor $S(q,\phi)$ 
according to 
\begin{equation} 
I(q,\phi) = \phi (\Delta \rho)^2 V_p P(q) S(q,\phi)\,, 
\label{S(q)} 
\end{equation} 
where $V_p$ is the volume of the solute per particle and 
$\Delta \rho = \bar \rho - \rho_m$ is the contrast of the 
solute resulting from the difference of the average scattering length density 
$\bar \rho$ and the scattering length density $\rho_m$ of the solvent (see Refs.~\cite{rose:06,Li:05} and further citations given there). From 
these definitions the volume fraction $\phi$ follows as $\phi = c \bar v$ 
where $c$ is the weight concentration of the dissolved polymer and $\bar v$ 
its specific volume in the respective solvent. The latter quantity can be obtained 
precisely from density measurements of dilute solutions 
 ($\bar v = 1.10 \pm 0.02$ cm$^3$/g).  
These data also serve for the calculation of $\Delta \rho = -5.67\times 10^{10}$ cm$^{-2}$. 
Figure \ref{fig2} displays SANS data obtained for various concentrations of the bottlebrush 
polymer dissolved in deuterated tetrahydrofurane. Additional investigations were done 
by static light scattering in order to explore the region of smaller $q$-values. 
These data have been used to obtain the molecular weight of the bottlebrush polymer.

For sufficiently small volume fractions $\phi$, the structure factor $S(q,\phi)$ 
can be expanded according to \cite{Li:05}
\begin{equation}
1/S(q,\phi) = 1 + 2 B_{app} \phi + O(\phi^2)\,,
\label{Bapp}
\end{equation}
where $B_{app}$ is the apparent second virial coefficient. Hence, Eq.~(\ref{Bapp}) 
suggests to plot $\phi/I(q,\phi)$ vs. $\phi$ for all 
$q$-values under consideration. The inset of Fig.~\ref{fig2} displays such a 
plot using the concentration $c$ instead of the volume fraction $\phi$. Straight lines 
are obtained allowing us to extrapolate the measured intensity to vanishing 
concentration. The open squares in Fig.~\ref{fig2} show the data obtained from this 
extrapolation together with the form factor obtained from the Pedersen-Schurtenberger 
\mbox{model 3} \cite{pede:96} which includes the effect of excluded volume 
(see also the discussion of this problem in Ref.~\cite{Zha:06}). The scattering intensity extrapolated 
to vanishing concentration is well-described by the model of the worm-like chain. 
We obtain the contour length \mbox{$L=380$ nm} and the persistence length 
\mbox{$l_p=17.5$ nm}. The radius of cross section of the chains follows as 5 nm. 
Static light scattering leads to a weight-average molecular weight of 
7.41 $\times 10^6$ g/Mol. Together with the weight-average degree of polymerization 
determined from the precursor polymer a molecular weight $M_0 = 4600$ g/Mol of 
the repeating unit results. Assuming a length of the repeating unit of 0.25 nm this 
would lead to a mass per unit length $M_L$ of 18.400 g/Mol/nm. Estimates of $M_L$ 
using the Holtzer plot (see, e.g., Ref.~\cite{Zha:06} and further literature given 
there) lead to a value of ca. 19.000 g/Mol/nm. Hence, the length of the repeating 
unit is ca. 0.24 nm which is slightly smaller than the calculated value of 0.25 nm. 
A similar finding was reported recently by Zhang and coworkers \cite{Zha:06}. 
Moreover, we have determined the contribution to the scattering intensity 
due to thermal fluctuations of the side chains.

\begin{figure}[t!]
\includegraphics[width=6.2cm,clip]{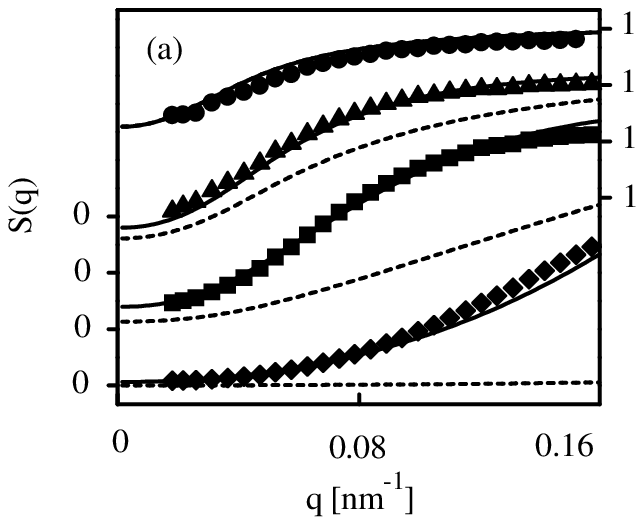}\\
\vspace*{0.5cm}
\includegraphics[width=6.0cm,clip]{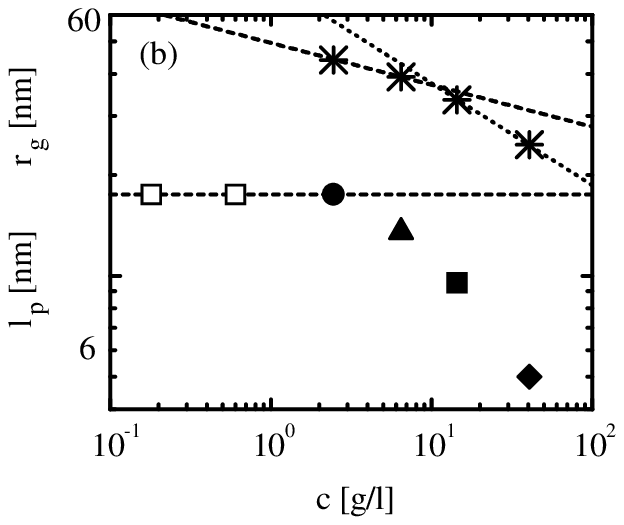}
\caption{(a) The structure factor $S(q,\phi)$ determined experimentally according 
to Eq.~(\ref{S(q)}) for four concentrations (with the same symbol code as in 
Fig.~\ref{fig2}). The dashed lines represent the structure factors as obtained 
from the PRISM integral equation theory [Eqs.~(\ref{eq3}) and (\ref{eq4})]
and assuming a {\it concentration-independent} shape of the bottlebrush polymers.
The solid lines represent the structure factors as obtained 
from the PRISM integral equation theory and using the {\it concentration-dependent} 
persistence lengths shown in (b) with the same symbol code (solid circle, triangle, 
square, and diamond). For reasons of clarity, the upper three data sets in (a) have 
been shifted up. For $c = 2.40$ g/l the dashed and solid curve coincide because the 
same persistence is used for both curves. The open squares in (b) denote two low 
concentrations which have been used for the extrapolation to infinite dilution 
as mentioned above. The radii of gyration as obtained from Eq.~(\ref{eq5})
are indicated in (b) by the stars. The upper dashed and dotted lines of slopes 
$c^{-1/8}$ and $c^{-17/56}$, respectively, represent two asymptotic scaling 
regimes \cite{bori:87}.}
\label{fig3}
\end{figure}

We now turn our attention to the analysis of the scattering intensities taken at 
finite concentration. The form factor $P(q)$ determined by extrapolating $I(q,\phi)$ 
to vanishing concentration is used to calculate $S(q,\phi)$ according to 
Eq.~(\ref{S(q)}). Fig.~\ref{fig3} (a) displays the experimental data  obtained 
for four different concentrations. 
A quantitative understanding of correlations and interactions between various 
colloidal and polymeric species can be achieved using the well-established 
techniques of liquid-state theory. The polymer reference interaction site 
model (PRISM) integral equation theory has been successfully applied to various 
systems, such as rodlike viruses \cite{harn:00}, plate-like colloids \cite{harn:01} 
and dendrimers \cite{rose:06}, flexible polymers \cite{harn:01a}, and mixtures of 
spherical colloids and semiflexible polymers \cite{harn:02}. Within the PRISM
theory the structure factor $S(q,\phi)$ reads 
\begin{equation} \label{eq3}
S(q,\phi)=1+\phi h(q,\phi)/(V_p P(q,\phi))\,,
\end{equation}
where $P(q,\phi)$ is the Fourier transform of the sum of the intramolecular 
two-point correlation functions for a given volume fraction $\phi$. In the 
limit $\phi\to 0$ this 
function reduces to the form factor $P(q)\equiv P(q,\phi\to 0)$. The total 
correlation function $h(q,\phi)$ describes correlations between different 
bottlebrush polymers, and is given by the generalized Ornstein-Zernike 
equation 
\begin{equation} \label{eq4}
h(q,\phi)=P^2(q,\phi)c(q,\phi)/(1-\phi c(q,\phi)P(q,\phi)/V_p)\,,
\end{equation}
where $c(q,\phi)$ is the  direct correlation function. This equation is 
solved numerically together with the Percus-Yevick closure taking steric 
interactions into account \cite{harn:01}. 

In Fig.~\ref{fig3} (a) the experimental structure factor $S(q,\phi)$ is compared 
to the results of the integral equation theory for the PRISM. We have used the form 
factor $P(q)$ (see the upper curve in Fig.~\ref{fig2}) as input into the generalized 
Ornstein-Zernike equation, i.e., $P(q,\phi)=P(q)$ in Eqs.~(\ref{eq3}) and (\ref{eq4}). 
With increasing bottlebrush polymer concentration the integral equation results 
(dashed lines) and the experimental data (symbols) deviate. The comparison of the 
calculated structure factors with the experimental data demonstrates that the 
{\it concentration-independent} persistence length $l_p=17.5$ nm and the form factor 
$P(q)$ may be used as input into the generalized Ornstein-Zernike equation only for 
very low concentrations of the bottlebrush polymers ($c \lesssim 2.5$ g/l). For 
higher concentrations marked deviations are found indicating that this approach 
is no longer valid. 

An alternative way of modeling these data is to consider a {\it concentration-dependent} persistence 
length of bottlebrush polymers and hence a concentration-dependent intramolecular 
correlation function $P(q,\phi)$ as input into Eqs.~(\ref{eq3}) and (\ref{eq4}). 
The results for the 
structure factors as obtained from the PRISM integral equation theory and using 
concentration-dependent persistence lengths are in agreement with the experimental 
data both for $S(q,\phi)$ (solid lines in Fig.~\ref{fig3} (a)) and $I(q,\phi)$
(four lower solid lines in Fig.~\ref{fig2}). The dependence of the persistence 
length on concentration shown in Fig.~\ref{fig3} (b) is reminiscent of the behavior 
of the predicted persistence length of polyelectrolytes (see Fig.~3 in Ref.~\cite{stev:93} 
and Fig.~4 in Ref.~\cite{shew:00}). Although the bottlebrush polymer solutions under 
consideration and the theoretically investigated polyelectrolyte solutions distinctly 
differ from each other, there is a significant overlap between them, namely the change 
of the shape of the polymers upon varying the concentration. Moreover, Fig.~\ref{fig3} (b) 
demonstrates that the concentration dependence of the calculated radii 
of gyration \cite{harn:96}
\begin{equation} \label{eq5}
r_g=\sqrt{Ll_p/3-l_p^2+2l_p^3/L-2(1-e^{-L/l_p})l_p^4/L^2}
\end{equation}
is in agreement with scaling considerations.

In addition to static properties we have investigated dynamic properties of the 
bottlebrush polymers using DLS. The measured time-dependent 
scattering intensity is a single exponential function of time for concentrations 
$c \lesssim 2.5$ g/l signaling pure translational diffusion of the polymers. 
No contributions of internal modes such a rotation, bending, or stretching to the 
dynamics have been found. We have determined the hydrodynamic radius
$R_h=39\pm 2$ nm from the measured translational diffusion coefficient. In order 
to understand the dynamic properties of the bottlebrush polymers at low concentrations 
it is instructive to compare the measured hydrodynamic radius with the results for
a semiflexible chain 
model which has been used to interpret quasi-elastic neutron and dynamic light 
scattering measurements on various natural and synthetic macromolecules 
\cite{harn:96} and worm-like micelles \cite{berl:98}. The numerical evaluation
yields \mbox{$R_h=38.5$} nm which is comparable with the experimentally determined value. 
Moreover, we have calculated the dynamic form factor and we have found that internal 
modes do not contribute to the dynamic form factor for the scattering vectors 
used in the light scattering experiments. However, internal modes do contribute 
for stiffer polymers confirming our findings concerning the stiffness 
of the bottlebrush polymers.

\begin{figure}[t!]
\includegraphics[width=6.2cm]{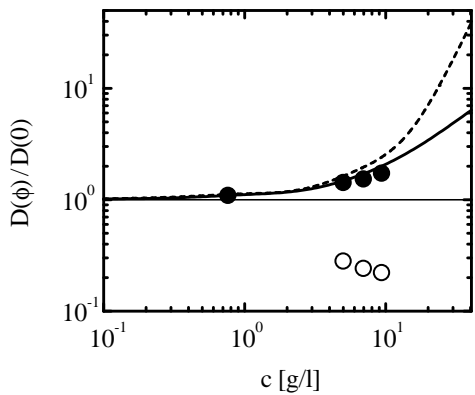}
\vspace*{-0.1cm}
\caption{Normalized cooperative diffusion coefficients $D(\phi)$ (solid circles) 
together with theoretical calculations using a concentration-independent [dependent]
shape of the bottlebrush polymers (dashed line) [(solid line)]. The open circles 
denote the measured diffusion coefficients of an additional slow diffusive 
process for concentrations $c \gtrsim 5$ g/l.}
\label{fig4}
\end{figure}

Figure \ref{fig4} demonstrates that the measured cooperative diffusion coefficient 
$D(\phi)$ (solid circles) increases upon increasing the bottlebrush polymer 
concentration due to an increasing restoring force for concentration fluctuations.
We have solved the equation 
$d I(q,\phi,t)/d t =-\Omega(q,\phi) I(q,\phi,t)$
for the time-dependent scattering intensity $I(q,\phi,t)$, where the decay rate 
$\Omega(q,\phi)$ depends on the solvent of the viscosity and the static scattering 
intensity $I(q,\phi)$ (see Ref.~\cite{harn:01a} for further details). As before in the 
case of the static correlation functions we have found that the cooperative diffusion 
coefficients $D(\phi)=\lim_{q\to 0} \Omega(q,\phi)/q^2$ as obtained from the 
equation for the time-dependent scattering intensity and using concentration-dependent 
persistence lengths (solid line in Fig.~\ref{fig4}) are in better agreement 
with the experimental data than the corresponding ones using the 
concentration-independent persistence length \mbox{$l_p=17.5$ nm} (dashed line in 
Fig.~\ref{fig4}). In addition we have observed experimentally a slow diffusive 
process at higher concentrations (open circles in Fig.~\ref{fig4}) which might be 
associated with long-range concentration fluctuations.

\begin{acknowledgments}
We wish to acknowledge financial support by the Deutsche Forschungsgemeinschaft, SFB 481, Bayreuth.
\end{acknowledgments}


\begin{thebibliography}{99} 

\bibitem{Win:94} M. Wintermantel {\it et al.}, 
Macromol. Chem. Rapid Comm. {\bf 15}, 279 (1994).

\bibitem{Win:95} M. Wintermantel {\it et al.}, 
Angew. Chem. Int. Ed. Engl. {\bf 34}, 1472 (1995).

\bibitem{Ger:99} M. Gerle {\it et al.}, Macromolecules {\bf 32}, 2629 (1999).

\bibitem{Zha:05} M. Zhang and A. H. E. M\"uller, J. Polym. Sci.: Polym. Chem. {\bf 43}, 3461 (2005).
\bibitem{Rat:05} S. Rathgeber {\it et al.}, 
J. Chem. Phys. {\bf 122}, 12904 (2005), and further references therein.

\bibitem{Rat:06} S. Rathgeber {\it et al.},
Polymer {\bf 47}, 7318 (2006).

\bibitem{Zha:06} B. Zhang, F. Gr\"ohn {\it et al.}, 
Macromolecules {\bf 39}, 8440 (2006).

\bibitem{Theory} S. Elli {\it et al.},
J. Chem. Phys.  {\bf 120}, 6257 (2004).

\bibitem{bori:87} O. V. Borisov, T. M. Birshtein, nad Y. B. Zhulina,
Polym. Sci. U.S.S.R {\bf 29}, 1413 (1987).

\bibitem{schw:92} K. S. Schweizer {\it et al.},
J. Chem. Phys. {\bf 96}, 3211 (1992).

\bibitem{yeth:92} A. Yethiraj and K. S. Schweizer, 
J. Chem. Phys. {\bf 97}, 1455 (1992).

\bibitem{mele:93} J. Melenkevitz {\it et al.},
Macromolecules {\bf 26}, 6190 (1993).

\bibitem{schw:97} K. S. Schweizer and J. G. Curro,
Adv. Chem. Phys. {\bf XCVIII}, 1 (1997).

\bibitem{sung:05} B. J. Sung and A. Yethiraj,
J. Chem. Phys. {\bf 122}, 234904 (2005).

\bibitem{stev:93} M. J. Stevens and K. Kremer,
Phys. Rev. Lett. {\bf 71}, 2228 (1993).

\bibitem{stev:95} M. J. Stevens and K. Kremer,
J. Chem. Phys. {\bf 103}, 1669 (1995).

\bibitem{yeth:97} A. Yethiraj,
Phys. Rev. Lett. {\bf 78}, 3789 (1997). 

\bibitem{yeth:98} A. Yethiraj,
J. Chem. Phys. {\bf 108}, 1184 (1998). 

\bibitem{shew:99} C.-Y. Shew and A. Yethiraj,
J. Chem. Phys. {\bf 110}, 5437 (1999)

\bibitem{shew:00} C.-Y. Shew and A. Yethiraj,
J. Chem. Phys. {\bf 113}, 8841 (2000). 

\bibitem{hofm:03} T. Hofmann {\it et al.},
J. Chem. Phys. {\bf 118}, 6624 (2003). 

\bibitem{Zha:03} M. Zhang {\it et al.}, 
Polymer {\bf 44}, 1449 (2003).

\bibitem{rose:06} S. Rosenfeldt {\it et al.}, 
ChemPhysChem {\bf 7}, 2097 (2006).

\bibitem{Li:05} L. Li {\it et al.}, 
Phys. Rev. {\bf 72}, 051504 (2005).

\bibitem{pede:96} S. Pedersen and P. Schurtenberger,
Macromolecules {\bf 29}, 7602, (1996)

\bibitem{harn:00} L. Harnau and P. Reineker, 
J. Chem. Phys. {\bf 112}, 437 (2000). 

\bibitem{harn:01} L. Harnau {\it et al.}, 
Europhys. Lett. {\bf 53}, 729 (2001).

\bibitem{harn:01a} L. Harnau, 
J. Chem. Phys. {\bf 115}, 1943 (2001). 

\bibitem{harn:02} L. Harnau and J.-P. Hansen, 
J. Chem. Phys. {\bf 116}, 9051 (2002). 

\bibitem{harn:96} L. Harnau {\it et al.},
J. Chem. Phys. {\bf 104}, 6355 (1996);
ibid. {\bf 109}, 5160 (1998);
Macromolecules. {\bf 30}, 6974 (1997); 
ibid. {\bf 32}, 5956 (1999).

\bibitem{berl:98} H. von Berlepsch {\it et al.}, 
J. Phys. Chem.  B. {\bf 102}, 7518 (1998). 











\end{thebibliography}
\end{document}